\documentstyle[prl,twocolumn,aps,epsf]{revtex}
\begin{document}
\twocolumn[\hsize\textwidth\columnwidth\hsize\csname@twocolumnfalse%
\endcsname
\title{Medium/high field magnetoconductance in chaotic quantum dots}

\author{E. Louis}

\address{Departamento de F{\'\i}sica Aplicada,
Universidad de Alicante, Apartado 99, E-03080 Alicante, Spain.}

\author{J. A. Verg\'es}

\address{{I}nstituto de Ciencia de Materiales de Madrid,
Consejo Superior de Investigaciones Cient\'{\i}ficas,
Cantoblanco, E-28049 Madrid, Spain.}

\date{\today}

\maketitle

\begin{abstract}
The magnetoconductance $G$ in chaotic quantum  dots at 
medium/high magnetic 
fluxes $\Phi$ is calculated by means of a tight binding Hamiltonian on a 
square lattice. Chaotic dots are simulated  by introducing diagonal 
disorder on surface sites of $L \times L$ clusters. 
It is shown that when the ratio $W/L$ is sufficiently large, $W$ being
the leads width, $G$ increases steadily showing a maximum
at a flux $\Phi_{\rm max} \propto W$. Bulk disordered ballistic
cavities (with an amount of impurities proportional to $L$) does not
show this effect. On the other hand, for magnetic fluxes larger than that 
for which the  cyclotron
radius is of the order of $L/2$,  the average magnetoconductance inceases
almost linearly with the flux with a slope proportional to $W^2$, shows
a maximum and then decreases stepwise. These results closely
follow a theory proposed by Beenakker and van Houten to  explain the 
magnetoconductance of two point contacts in series. 
\end{abstract}

]

\narrowtext
\section{Introduction}
Magnetoconductance in chaotic quantum dots have attracted a great
deal of attention in recent years \cite{MR92,TN97,GM98,SK98}.
Weak localization effects  have been thoroughly investigated 
searching for differences between chaotic and regular cavities
\cite{MR92,GM98}. More recently the selfsimilar character of
magnetoconductance fluctuations in chaotic quantum dots has
deserved several experimental studies \cite{TN97,SK98}. In
Ref. \cite{SK98} it was reported that in cavities with sufficiently 
soft walls, rather wide leads and a high zero field conductance
of the order of 40 conductance quanta, fluctuations were very weak and 
the magnetoconductance
increased steadily in approximately  20\% over 50 flux quanta \cite{SK98}. 
Although an increase in the magnetoconductance as a function of the 
magnetic flux in cavities with wide leads 
may not be that surprising, no theoretical analysis of this result  
is yet available. This is so despite of the wealth of experimental
and theoretical information on magnetoconductance in
related mesoscopic systems \cite{BH91,Da95,AA82,Be97,SM90,BH89}. 

The purpose of the present work is to investigate the magnetoconductance
of chaotic quantum dots over a wide range of magnetic field. 
Quantum dots are described by means of a tight--binding Hamiltonian on $L
\times L$ clusters of the square lattice. Non--regular (chaotic) behavior 
is induced by introducing disorder at surface sites, a procedure that
has been shown to reproduce all properties of quantum chaotic cavities
\cite{CL96}. The most outstanding conclusions derived from our study are the
following. For sufficiently open systems, large leads width $W$ or, 
alternatively, high zero field conductance, the magnetoconductance
increases steadily as a function of the magnetic flux, reaching a maximum 
at a magnetic flux $\Phi_{max}$ proportional to the the leads width. This
effect, which is in agreement with the experimental observations of \cite{SK98},
does not show up in regular or bulk disordered cavities.
The average magnetoconductance versus magnetic flux curve shows four clearly 
differentiated regions:  i) At small fluxes (typically below 1--2 flux
quanta) the weak localization peak with the typical Lorentzian shape
is observed. ii) This is followed by a flux range over which the 
magnetoconductance shows a non--universal 
behavior which depends on leads configuration. iii) The latter
lasts until the cyclotron radius becomes of the order of $L/2$. Beyond
this point the average magnetoconductance increases linearly with the magnetic
flux with a slope which increases with the leads width (for
very small $W$ the slope is nearly zero and the magnetoconductance remains 
constant in a large flux range) . iv) At large fluxes the magnetoconductance 
decreases stepwise (each step of one flux quanta), due to the 
successive crossings of the Fermi energy of the transversal modes 
that contribute to the current. As discussed below these results are compatible
with a theory proposed by Beenakker and van Houten \cite{BH89} to
interpret the experimental results for the magnetoconductance of two point 
contacts in series \cite{SM90}.

The rest of the paper is organized as follows. Section II includes a 
description of our model of chaotic quantum dot and of the method we used
to compute the current. The results are presented in Section III, and
discussed in terms of the theory of Ref. \cite{BH89}  in Section III.
Section IV is devoted to summarize the conclusions of our work.

\section{Model and Methods}

\subsection{Model}
The quantum dot is described by means of a
tight-binding Hamiltonian with a single atomic level per lattice site
on $L \times L$ clusters of the square lattice:

\begin{eqnarray}
\widehat H&=&\sum_{m,n~\epsilon ~ IS} \omega_{m,n} |m,n><m,n| \nonumber \\ 
&&- \sum_{<m,n;m',n'>} t_{m,n;m',n'}|m,n><m',n'|,
\end{eqnarray}

\noindent
where $|m,n>$ represents an atomic orbital on site $(m,n)$.
Indexes run from 1 to $L$, and the symbol $<>$ denotes that the sum
is restricted to nearest-neighbors. Using Landau's gauge the hopping 
integral is given by,
\begin{eqnarray}
t_{m,n;m',n'}&=&{\rm exp}\left (2\pi i \frac{m}{(L-1)^2}\Phi
\right ), ~~ m=m' \nonumber \\
&&=~1,~~~~~{\rm otherwise} 
\end{eqnarray} 
\noindent where the magnetic flux $\Phi$ is measured in units of the 
quantum of magnetic flux  $\Phi_0 = h/e$.
The energy $\omega_{m,n}$ of atomic levels at impurity sites ($IS$)
is randomly chosen between $-\Delta/2$ and $\Delta/2$, whereas at other sites
$\omega_{m,n} = 0$. Impurities were  taken on all surface sites 
\cite{CL96,BM98} but those coinciding with the leads entrance sites
to avoid excessive (unphysical) scattering. This model has been proposed
to simulate cavities with rough boundaries and its properties closely 
follow those characterizing quantum chaotic systems
\cite{CL96,BM98}. Some calculations were also carried out on clusters
with $2L$ bulk impurities \cite{note1}.

\subsection{Conductance}
The conductance
(measured in units of the quantum of conductance $G_0 = e^2/h$) was
computed by using an efficient implementation of Kubo formula. The
method is described in \cite{Ve99}, while applications to mesoscopic systems 
can be found in \cite{CL97,LV00}. 
For a current propagating in the $x$--direction,
the static electrical conductivity is given by:

\begin{equation}
G = -2 {{\left( \frac{e^2}{h} \right)} {{\rm Tr} \left [(\hbar {\hat v_x})
{\rm Im\,}{\mathcal \widehat G}(E)(\hbar {\hat v_x})
{\rm Im\,}{\mathcal \widehat G}(E)\right ]}} \;,
\end{equation}
where ${\rm Im\,}{\mathcal \widehat G}(E)$ is obtained from
the advanced and retarded Green functions:
\begin{equation}
{\rm Im\,}{\mathcal \widehat G}(E)=\frac{1}{2i}\left[{\mathcal
\widehat G}^{R}(E)-{\mathcal \widehat G}^{A}(E)\right ] \;,
\label{e:img}
\end{equation}
and the velocity (current) operator ${\hat v_x}$ is related
to the position operator ${\hat x}$ through the equation of motion
$\hbar {\hat v}_x = \left [ {\widehat H},{\hat x} \right ]$,
$\widehat{H}$ being the Hamiltonian.

Numerical calculations were carried out connecting quantum dots
to semiinfinite leads of width W in the range 1--$L$.
The hopping integral inside the leads and between leads and dot
at the contact sites is taken equal to that in the
quantum dot (ballistic case). Assuming the validity of both
the one-electron approximation and linear response,
the exact form of the electric field does not change the value of $G$.
An abrupt
potential drop at one of the two junctions provides the simplest
numerical implementation of the Kubo formula \cite{Ve99} since, in this case,
the velocity operator has finite matrix elements on only two adjacent
layers and Green functions are just needed for this restricted
subset of sites. Assuming this potential drop to occur at the left contact
($lc$) side, the velocity operator can be explicitly written as,
\begin{equation}
i\hbar v_x = -\sum_{j=1}^W\left (|lc,j><1,j|-|1,j><lc,j|\right)
\end{equation}
\noindent where $(|lc,j>$ are the atomic orbitals at the left contact sites
nearest neighbors to the dot.

Green functions are given by:
\begin{equation}
[E \widehat I - \widehat H - \widehat \Sigma_{\mathrm 1}(E) - \widehat
\Sigma_{\mathrm 2}(E)] {\mathcal \widehat G}(E) = \widehat I \;,
\label{e:green1}
\end{equation}
where $\widehat \Sigma_{1 , 2}(E)$
are the selfenergies introduced by the two semiinfinite leads \cite{Da95}.
The explicit form of the retarded selfenergy due to the mode of
wavevector $k_y$ is:
\begin{equation}
\Sigma(E)={{1 \over 2} \left( E -\epsilon(k_y)- i
\sqrt {4-(E-\epsilon(k_y))^2} \right) } \;,
\label{e:sigma}
\end{equation}
for energies within its band $|E -\epsilon(k_y)| < 2$, where
$\epsilon(k_y)=2{\rm cos}(k_y)$ is
the eigenenergy of the mode $k_y$ which is quantized as
$k_y=(n_{k_y}\pi)/(W+1)$,
$n_{k_y}$ being an integer from 1 to $W$. The transformation from the normal
modes
to the local tight--binding basis is obtained from the amplitudes of the
normal modes, $<n|k_y> = \sqrt{2/(W+1)}{\rm sin}(nk_y)$.
Note that in writing Eq.~(\ref{e:sigma}) we assumed that the magnetic field
was zero outside the dot \cite{Da95}. This point will have some relevance 
in relation to the interpretation of our numerical results in terms of
the theory of Ref. \cite{BH89}.

Calculations were carried out on clusters of linear
size $L=47-394$ (in units of the lattice constant) and at a fixed  
arbitrarily chosen, Fermi energy $E=-\pi/3$. In some cases averages 
over disorder realizations were also done. 
Although most calculations were performed with input/output leads 
of width $W$  connected from site $(1,1)$ to site $(1,1+W)$, and 
from $(L,L)$ to $(L,L-W)$, respectively, other input/output leads 
configurations were also explored. All calculations on disordered
systems were done at a fixed value of the disorder parameter $\Delta$=6.

\section{Results}

We first discuss results for a $W/L$ similar to that used in the experiments
of Ref. \cite{SK98}. In that work conductance measurements were taken on
a stadium cavity with a lithographic radius of 1.1 $\mu$m and leads
0.7 $\mu$m wide, which gives a $W/L$ ratio of 0.64.
Fig.~\ref{f:peak} shows the results for the magnetoconductance in 
cavities of linear sizes in the range $L$=47--394 and leads of width $W=0.65L$. 
The results correspond to a single realization of disorder. The most 
interesting result is the steady increase of the conductance with the
magnetic field. The conductance reaches a maximum at a magnetic
flux which, as illustrated in the inset of the Figure, increases linearly
with the leads width, or the linear size of the system.  
Note that, as the results of Fig.~\ref{f:peak} correspond to $W \propto L$,
they cannot allow to identify which of the two parameters ($W$ and $L$)
control the maximum in $G$. We have checked, however, that it is in fact $W$
the one that matters (see also below).
The increase in $G$ until the maximum is reached can be as high as 30\%. 
It is interesting to remark that the increase in the conductance 
occurs with relatively small fluctuations due to the large $W/L$ ratio 
(or degree of opening) of the cavity (see \cite{LV00}). Although the 
experimental data were taken at fields not high enough to observe the
maximum shown in Fig.~\ref{f:peak}, it can be safely assessed that our
results are compatible with those of Ref. \cite{SK98}. In particular 
the zero field conductance reported in that work was around 38 quanta,
increasing up to 46 quanta over approximately 50 flux quanta. This is rather
similar to the magnetoconductance curve for $L=197$ shown in Fig.~\ref{f:peak}.

In Fig.~\ref{f:bulk} we compare the results for the cavity
with surface disorder with those for a regular cavity and for a cavity with
$2L$ bulk impurities. The results indicate that the cavity
having surface disorder is the only one that reproduces the experimental
results \cite{SK98}. In regular cavities the conductance
does not increase steadily due to the large amplitude 
oscillation discussed in \cite{LV00}. This behavior clearly differentiates 
regular and chaotic cavities. At large fields the result for the cavity with 
surface disorder coincides with that for the regular cavity. This is
a consequence of the fact that for sufficiently high fields the current is
dominated by edge--like states which are not affected by surface disorder.
Semiclassically one can view carriers motion as short orbits bouncing off the
same boundary. The associated quantum states  have chirality and 
are thus commonly refer to as chiral states or edge states. 
The stepwise decrease of the magnetoconductance observed in regular and 
chaotic cavities with surface disorder is a consequence of the overall
depopulation of Landau levels.  It is interesting to note that 
the cavity with bulk disorder
shows a markedly different behavior as this type of disorder can, instead, 
scatter carriers between opposite sides of the cavity. The results shown in 
Fig.~\ref{f:bulk} illustrate the only difference we have found up to now 
between cavities with surface impurities or with a number of bulk impurities
proportional to $L$ \cite{VL99}. Appart from this difference the two behave 
much alike and in line with what one expects to be the behavior of 
quantum chaotic cavities \cite{GM98}. 

We have investigated how this steady increase of the magnetoconductance 
is affected by the leads width. In order to reduce fluctuations, which are 
particularly important at small $W$ \cite{LV00}, we have averaged 
the conductance over disorder realizations (600 realizations were included in 
the calculations). 
The results for a cavity with rather narrow leads attached at the
dots in three different ways are illustrated in Fig.~\ref{f:widerange}.  
At small fluxes (typically below 1--2 flux quanta) the expected Lorentzian 
peak characteristic
of chaotic cavities \cite{GM98} is obtained (not clearly visible in 
the Figure). At higher fluxes, a range over which
the conductance behaves in a way that is strongly dependent on leads
configuration, is observed. Beyond, the conductance increases linearly 
with a slope  which is very similar in the three cases shown in the Figure.
The crossover to the linear behavior occurs at a flux for
which the radius of the classical cyclotron orbit $r_c$ is  roughly
$L/2$. In our units (flux and conductance quanta $\Phi_0=G_0=1$ and
energy $\hbar^2/(2ma^2)=t=1$, where $t$ is the hopping integral), 
$r_c$ is given by,
\begin{equation}
r_c = \hbar v(E) \frac{(L-1)^2}{4\pi}\frac{\Phi_0}{\Phi}\;,
\end{equation}
\noindent where $\hbar v(E) = \langle 2\sqrt{{\rm sin}^2k_x+{\rm sin}^2k_y}
\rangle_E$. At the energy chosen here $\hbar v(E) \approx 2.2$ \cite{CL97}.
Then the flux at which $r_c = L/2$ is, for the size of Fig.~\ref{f:widerange},
$\Phi = 17\Phi_0$, which is very close to the flux at which the mentioned
crossover occurs. 

Fig.~\ref{f:linear} shows the  averaged magnetoconductance for cavities of 
linear size $L=47$ and leads widths in the range $W$=4--20. It is noted that
the linear behavior discussed above appears in all cases at roughly the
same magnetic flux indicating that it is only related to the
cavity size $L$, as suggested by the discussion above. Instead the slope 
increases with the leads width. At small $W$ (systems with a low 
conductance) the conductance increases very slowly as a function of the 
magnetic flux. For sufficiently large $W$ the conductance reaches a maximum
and then decreases stepwise. As remarked above, the latter is a consequence 
of the overall depopulation of transversal modes (or Landau levels).
The slope of the linear part of the 
magnetoconductance is plotted as a function of leads width in 
Fig.~\ref{f:slope}. The results can be accurately fitted by a $W^2$ law. 
The increase of the magnetoconductance can be understood in terms of
the increase of the transmission probability of the transversal modes
as their edge--like character increases and, consequently, its 
sensitivity to surface disorder is reduced. 
The steady increase in the conductance takes place until
the mentioned depopulation begins to reduce the number of modes that 
participate 
in the current. Although this argument seems plausible, it cannot explain
quantitative features of the results such as  the linear relation 
between the conductance and the flux or the increase of the 
slope as the square of the leads width. This issue is addressed in the
following Section.

\section{Discussion}

The results discussed in the previous subsection ressemble those predicted
by Beenakker and van Houten for the magnetoconductance of two point
contact in series \cite{BH89} and the related experiments of
Staring {\it et al} \cite{SM90}.  Under the hypothesis that transmission
between point contacts occurs with intervening equilibration of the
current--carrying edge states, the authors of \cite{BH89} derived
the following expression for the conductance (in the
following we take the conductance and flux quanta $G_0 = \Phi_0 = 1$ 
and do not include spin degeneracy),
\begin{equation}
G(\Phi)=\left [ \frac{1}{N_1}+\frac{1}{N_2}-\frac{1}{N_L}\right ] ^{-1}
\label{e:G_BH}
\end{equation}

\noindent where $N_i$ are the number of occupied subbands in the two
contacts or leads ($i=1,2$) and in the region between the contacts,
or in the present case in the dot ($i=L$). Disregarding 
discretness \cite{BH89}, $N_i$ can be written as,
 
\begin{equation} 
N_i=\frac{n_i}{2B}f(\xi_i)
\label{e:ni}
\end{equation}
\noindent where the function $f(\xi_i)$ is,

\begin{eqnarray}
f(\xi_i)&=&\frac{2}{\pi}\left[{\rm arcsin}\xi_i+\xi_i(1-\xi_i^2)^{1/2}
\right],\;{\rm if}\;\; \xi_i < 1 \nonumber \\
&&=~1,~~ \; {\rm if}\;\; \xi_i>1
\end{eqnarray}
\noindent with $\xi_i=l_i/2r_c$, $l_i$ being a characteristic linear 
dimension in the three regions, in the present case, $l_i = W_1, W_2, L$.

We have used Eqs.~(\ref{e:G_BH}) and (\ref{e:ni}) to fit the numerical results
of Fig.~\ref{f:linear}. We took as fitting parameter 
the density or the Fermi velocity, $\hbar v=2\sqrt{2\pi n}$, and assumed
the same density in the leads and dot. As shown in Fig.~\ref{f:theory}
a satisfactory fitting is obtained for $\hbar v =  3.65$, almost twice 
the actual Fermi velocity in our model (see above). The theory
reproduces the three regions that characterize our numerical results:
an almost constant $G$ for small flux or $r_c > L/2$, a steadily
increasing $G$  up to $r_c \approx W/2$ followed by a steep decrease
at higher fluxes. It is interesting to note that the theory of Ref.\cite{BH89}
show a better agreement
with the numerical results for the case in which the leads are attached
at opposite corners of the dot, than with those for the other two
lead configurations of Fig.~\ref{f:widerange}. A possible reason for this 
behavior relies upon the equilibration assumption
in Beenhakker and van Houten theory. Equilibration is most likely when the 
leads are not facing each other, as is the case of leads attached to
opposite corners. Instead, when leads are attached at contiguous corners
direct transmission is more probable and equilibration requires higher
magnetic fields to take place. This is a pictorial illustration of the
assumptions under which the theory of Ref. \cite{BH89} holds.

In order to check whether a linear relationship between the conductance
and the flux in a rather wide range of fluxes, as indicated
by the fittings of Fig.\ref{f:linear}, can be understood in
terms of this theory, we have expanded the conductance for small $\xi$ 
and $\xi_L > 1$. The result for leads having the same width
and the same Fermi velocity (or density) in the leads and dot, is
\begin{equation}
G(\Phi) \approx \frac{n_1}{2\pi B}\left[2\xi + \frac{4}{\pi}\xi^2+
\left(\frac{8}{\pi^2}-\frac{1}{3}\right)\xi^3\right]
\label{e:expansion1}
\end{equation}
\noindent As the maximum in $G$ occurs for $\xi$ slightly smaller than 
unity, checking whether $G$ varies linearly with the magnetic field 
below the maximum requires only to calculate the ratio between the
coefficients of the second and third power of $\xi$ in Eq.(\ref{e:expansion1}). 
This ratio is 2.82, indicating that the linear term dominates in agreement 
with our numerical results. This equation
also shows that the slope of the straight line is proportional to
$W^2$. To make a quantitative comparison with the result of
Fig.~\ref{f:slope} we rewrite Eq.~(\ref{e:expansion1}) introducing the
actual expression for $\xi$; the result is,

\begin{equation}
G(\Phi) \approx \frac{\hbar v}{4\pi}W + \frac{W^2}{\pi L^2}\Phi+
\frac{\pi}{2\hbar v}\left(\frac{8}{\pi^2}-\frac{1}{3}\right)
\frac{W^3}{L^2}\Phi^2
\label{e:expansion2}
\end{equation}

\noindent The coefficient of the linear term results to be $1.44 \times
10^{-4} W^2$ not too far from the numerical result of Fig.~\ref{f:slope}.

Before ending it is worth to comment on several differences between
our model calculation and the theory of Ref. \cite{BH89}. We first note that we 
have assumed that the magnetic field is zero outside the dot, which may
at first sight invalidate the use of Eq.~({\ref{e:ni}). Nevertheless, the 
Green function of the whole system is calculated through Dyson's 
equation (see Eq.~(\ref{e:green1}). This means that the region of the leads 
close to the dot is distorted by the magnetic field, which seems to 
be enough
to validate the calculation of the number of occupied subbands
by means of Eq.~({\ref{e:ni}). On the other hand whereas in the theory
of Ref.~\cite{BH89}) each channel contributes with one quanta
to the conductance, in our case this contribution is approximately halved
(remember that we work on chaotic cavities, see Ref.~\cite{GM98}).
This seems, however, to be irrelevant as far as the qualitative behavior 
of the conductance is concerned.

\section{Concluding Remarks}
Summarizing, we have presented a numerical analysis of the magnetoconductance
of quantum chaotic cavities in a wide range of magnetic fields. For
sufficiently open cavities the magnetoconductance increases steadily
reaching a maximum at a flux proportional to the leads width.
This steady increase of $G$ agrees with the experimental observations
reported in \cite{SK98}.  Neither
regular nor bulk disordered cavities behave in this way. Numerical results 
for the average magnetoconductance indicate that, for magnetic fluxes larger 
than that for which the  cyclotron
radius is approximately $L/2$ and smaller than the flux at which the
mentioned maximum is reached, it increases linearly with the magnetic flux
$\Phi$ with a slope  proportional to the square of the leads width. At
higher fluxes the conductance decreases stepwise. These results admit a
satisfactory explanation in terms of the theory proposed by Beenhakker
and van Houten to interpret the experimental
results for the magnetoconductance of two contacts in series. 
The fact that our results for small magnetic fluxes ($r_c > L/2$)
show a better agreement with the theory in the case that the two contacts 
are attached to opposite corners of the dot (and, thus, are not facing 
each other), 
is related to the stronger equilibration of edge--states
promoted by this lead configuration with respect to the other two 
geometries explored in this work.

\acknowledgments
We thank C.W.J. Beenakker and D. Khmelnitskii for very
useful correspondence, and L. Brey, C.Tejedor and J. Palacios 
for some interesting comments and remarks.
This work was supported in part by the Spanish CICYT (grants PB96-0085 and
1FD97-1358).

\begin{figure}
\begin{picture}(236,190) (-10,-15)
\epsfbox{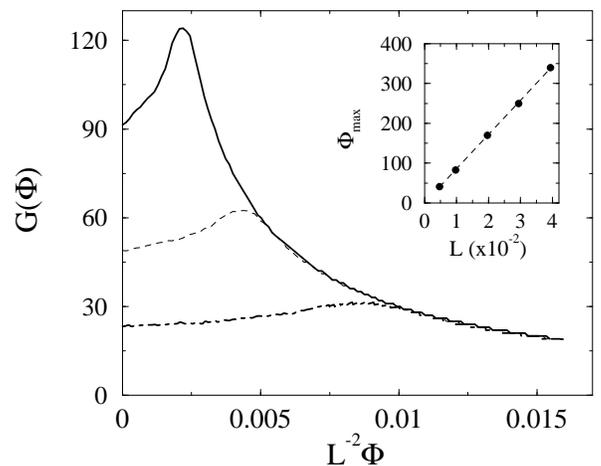}
\end{picture}
\caption{
Magnetoconductance versus magnetic flux multiplied by the inverse
of the dot area (both in units of their respective quanta), in dots of linear 
size $L$ and leads of
width $W$ (connected at opposite corners) with a similar $W/L$ 
ratio of $\approx 0.65$. 
Results for $(L,W)$=(97,63) chain line, (197,127) broken line and (394,254) 
continuous line, are shown. The results correspond to a single realization
of disorder and Fermi energy $E=-\pi/3$. Inset: the flux at
which the magnetoconductance is maximum is plotted as a function of the linear
size of the dot $L$. The fitted straight line is, $\Phi_{max}/\Phi_0=
0.026+0.86L$.
\label{f:peak}}
\end{figure}

\begin{figure}
\begin{picture}(236,175) (-10,-10)
\epsfbox{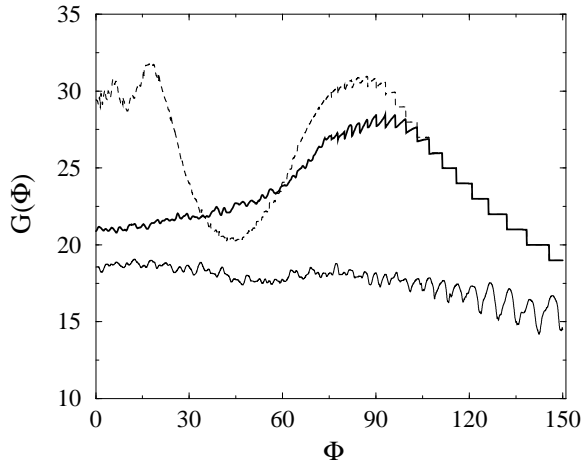}
\end{picture}
\caption{
Magnetoconductance versus magnetic flux (both in units of their
respective quanta) in $97 \times 97$ dots with leads of width $W=57$
connected at opposite corners of the dot. The results correspond to
dots with: i) no disorder (broken line), ii) Anderson impurities
with $\Delta$=6 (a single realization of disorder) placed either
on all surface sites but those coinciding with the leads entrance sites
(continuous line) or  $2L$ impurities distributed randomly within the 
dot (thin continuous line).
\label{f:bulk}}
\end{figure}

\begin{figure}
\begin{picture}(236,190) (-10,-10)
\epsfbox{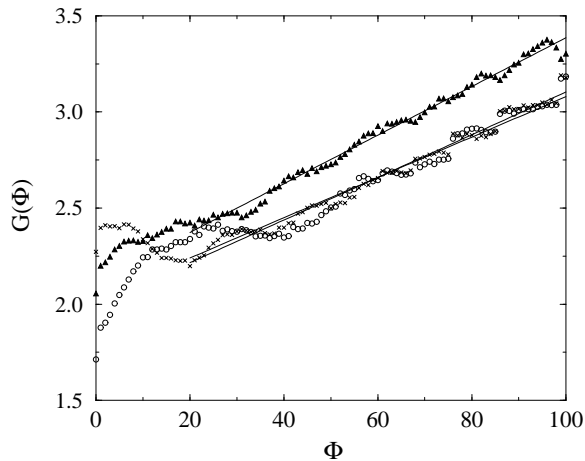}
\end{picture}
\caption{
Magnetoconductance  versus magnetic flux 
in cavities of linear size $L=47$ and leads of width 8 connected at:  
contiguous corners (circles), opposite corners (crosses), and, one corner 
and the center of the opposite side (triangles). 
Anderson impurities with $\Delta=6$
were placed at all surface sites but those coinciding with the leads
entrance sites. The results correspond to an energy $E=-\pi/3$ and an average
over 600 disorder realizations. Straight lines were fitted for fluxes above 
20. The fitted lines are: crosses  $G=1.99+0.011\Phi$, circles 
$G=2.03+0.01\Phi$, and triangles $G=2.12+0.013\Phi$. 
\label{f:widerange}}
\end{figure}

\begin{figure}
\begin{picture}(236,275) (-10,-0)
\epsfbox{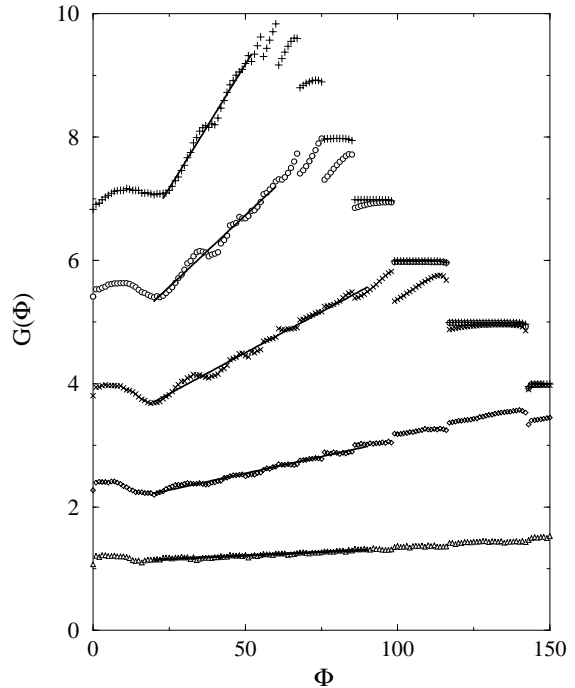}
\end{picture}
\caption{
Magnetoconductance  versus magnetic flux 
in cavities of linear size $L=47$ and leads of width  $W$ = 4 (triangles), 
8 (diamonds), 12 (crosses), 16 (circles) and 20 (+), connected at opposite 
corners of the dot. Anderson impurities with $\Delta=6$
were placed at all surface sites but those coinciding with the leads
entrance sites. The results correspond to an energy $E=-\pi/3$ and an average
over 600 disorder realizations. Straight lines were fitted for fluxes above 
20 and below the flux at which the conductance shows a maximum. The 
slopes of the straight lines are plotted as a function of $W^2$ in 
Fig.~{\protect \ref{f:slope}}.
\label{f:linear}}
\end{figure}

\begin{figure}
\begin{picture}(236,170) (-20,-15)
\epsfbox{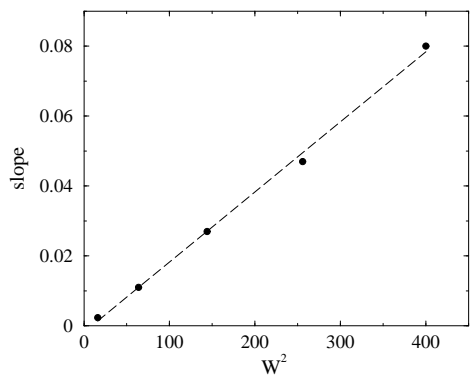}
\end{picture}
\caption{
Slopes of the lines fitted in Fig.~{\protect \ref{f:linear}} versus
the square of the leads width $W$. The fitted straight line  is 
$-0.0019+2\times10^{-4}W^{2}$.
\label{f:slope}}
\end{figure}

\begin{figure}
\begin{picture}(236,200) (-20,-15)
\epsfbox{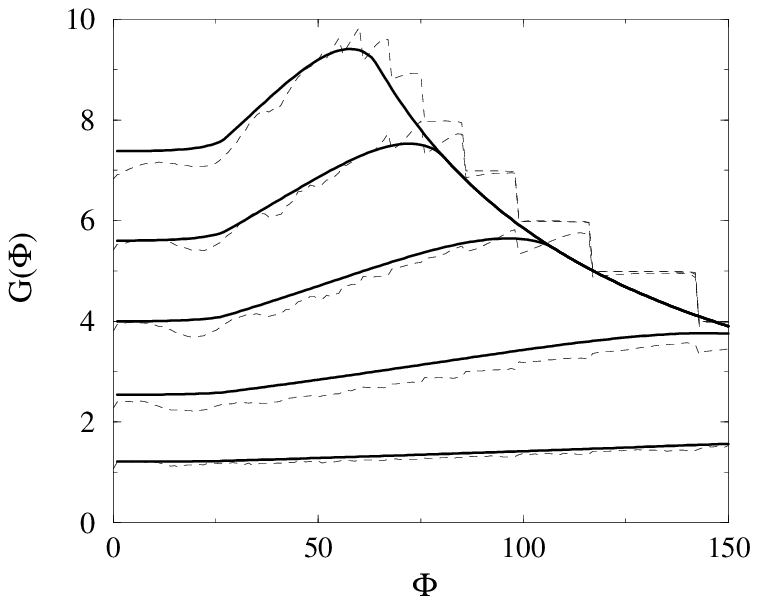}
\end{picture}
\caption{
Fitting of the numerical results of Fig.~{\protect \ref{f:linear}}
by means of the theory of Ref. {\protect \cite{BH89}}, 
broken and continuous curves respectively (see text).
\label{f:theory}}
\end{figure}


\begin{references}
\bibitem{note1}
This model has been shown to follow the characteristics
of quantum chaotic systems {\protect \cite{VL99}}. The key point is that
chaotic behavior is obtained whenever the number of impurities is
proportional to $L$.
\end{references}
\end{document}